\documentclass[12pt,a4paper]{article}
\usepackage{amssymb,latexsym}
\usepackage{amsmath}
\usepackage[dvips]{graphicx}

\title{\textbf{Composite structure of vortices in two-component Bose-Einstein condensate}}
\author{\textbf{Anatoly P. Ivashin\footnote{ivashin@kipt.kharkiv.ua} ,  Yuri M. Poluektov \footnote{yuripoluektov@kipt.kharkov.ua}}\\
 National Science Center ``Kharkiv Institute of Physics
and Technology''\\
\textit{Akademichna Str.,1, Kharkov, 61108, Ukraine}}

\date{}

\begin{document}

\maketitle

\begin{abstract}	

In contrast to one-component Bose-Einstein condensate case,
the vortices in two-component condensate can have various complicated structures.
The vortices in a space-homogeneous Bose-Einstein condensate have been studied in this paper.
It is shown that the vortex structure is described by three dimensionless parameters. 
This is totally different from the usual one-component condensate case,
where an isolated vortex is described by a parameter-less dimensionless equation.
The two-component vortex structure strongly depends on the sign of ``interaction'' constant of the components.
A few types of vortices with different qualitative structure are explored.
We show that the super-density vortices can exist, when the ``interaction'' constant is positive.
The super-density vortices have the near-axis density greater than the equilibrium density of a homogeneous space Bose-Einstein condensate.
We also show that the vortices with opposite direction of the condensate component rotation near the axis and far off the axis
can exist.

\end{abstract}
\begin{flushleft}
\end{flushleft}

%Ser'ga
%\documentclass{article}
%\usepackage[english,russian]{babel}
%\usepackage{amssymb}
%\usepackage{amsmath}
%\usepackage[dvips]{graphicx}

\textit{}

 In~\cite{P,G} the quantized vortex filament in a weakly nonideal Bose gas was investigated. The structure of the vortex is on the axis of the vortex the following density of a Bose-Einstein condensate (BEC) vanishes and increases monotonically to the equilibrium value of the density of a spatially homogeneous system at infinity. The characteristic distance over which the density varies near the vortex axis is determined by the correlation length, which depends on the mass of the particle, the equilibrium density and interparticle interaction constant. The correlation length determines the size of the vortex core. The equation describing the vortex in a single-component system, if it is written in dimensionless form, is universal and does not contain any parameters, characterizing the system~\cite{LP}. Consequently, the structure of the vortices in all single-component systems is the same. The influence of nonlocal effects on the vortex structure was seen in~\cite{IP}.

Creating a BEC in atomic gases of alkaline elements contained in magnetic or laser traps~\cite{P2,KAD,PethSm,PStr}, led to experimental study of the vortices in spatially inhomogeneous conditions~\cite{LGCK,SSV,MBG,SAC,MAHH}. Were examined not only single-component systems, but has been implemented the ability to create condensates mixture of two or more types of particles and vortex states in these more complex systems~\cite{MBG,SAC, MAHH}. Although the two-component BEC in trap with vortices have been studied theoretically~\cite{PethSm,PStr,Mach,HO,HO2,GRPG,FS,KTU}, a detailed analysis of the structure of the vortex in such systems, depending on the characteristics of the environment, as we know, is absent for the spatially homogeneous case.

In this paper we study the structure of vortices in the spatially homogeneous Bose-Einstein condensate of two kinds of particles with different masses. It is shown that in this two-component case a  structure of the vortex depends on the three-dimensional parameters characterizing the system. Changes in the density of each component, the total density and mass flux density as a function of distance from the axis of the vortex is determined by the values of these parameters. Attention is drawn that the considered system differs significantly from the one-component case where the vortex is described by a universal nonlinear equation in dimensionless form which does not contain the characteristics of the system~\cite{P,G,LP}.
 Some possible qualitatively different vortex structure with different sets of parameters characterizing the two-component condensate is explored. It is shown that the structure of the vortex depends on the sign of  ``interaction'' component constant.
In particular, the existence of a sort of ``super-density vortices'' is possible in case with positive value of the coupling constant. In such vortices total density near the axis may exceed the equilibrium density spatially homogeneous condensate. This is impossible in single-component case. It is also shown that there are vortices in which the condensate near the axis and at large distances from the axis of the vortex is rotated in opposite directions.

In the first section of the paper the system of the Gross-Pitaevskii (GP) equations for a two-component BEC is presented in dimensionless symmetrical form. The system contains five dimensionless parameters, among  them only three parameters are independent.
In the second section the GP equations for macroscopic wave functions absolute values are written in the cylindrical coordinate system in the presence of vortex motion with two quantized circulation. In the third section we investigate the asymptotics of solutions at large distances from the vortex axis. It is shown that the solution of the equations may tend to its asymptotic value from above and below, thus under certain conditions the total number density near the axis of the vortex can be greater than the equilibrium density at infinity.
Some qualitatively different vortex structures are numerically studied in the fourth section. We consider the cases when the circulation is different from zero in one of the components, or in both of them. In the fifth section we analyze the distribution of the mass flow density in the vortex filaments.

%%%%%%%%%%%%%%%%%%%%%%%%%%%%%%%%%%%%%%%%%%%%%%%%%%%%%%%%%%%%%%%%%%%%%%%%%%%%%%%%%%%%%%%%%%%%%%%%%%%%%%

\section{Gross-Pitaevskii equation for two-component system}

%%%%%%%%%%%%%%%%%%%%%%%%%%%%%%%%%%%%%%%%%%%%%%%%%%%%%%%%%%%%%%%%%%%%%%%%%%%%%%%%%%%%%%%%%%%%%%%%%%%%%%%

Bose Einstein condensate, consisting of two particle species with mass $m_{1} $, $m_{2}$ and densities $n_{1} ,\; n_{2} $ is described by stationary Gross-Pitaevskii equations for macroscopic wave functions
 
\begin{eqnarray} 
\label{eq1a}
-\frac{\hbar ^{2} }{2m_{1} } \Delta \Phi _{1} \left(\mathbf{r}\right)+g_{11} \Phi _{1} \left(\mathbf{r}\right)\left|\Phi _{1} \left(\mathbf{r}\right)\right|^{2} +g_{12} \Phi _{1} \left(\mathbf{r}\right)\left|\Phi _{2} \left(\mathbf{r}\right)\right|^{2} 
&=&
\mu _{1} \Phi _{1} \left(\mathbf{r}\right),
\\ 
\label{eq1b}
-\frac{\hbar ^{2} }{2m_{2} } \Delta \Phi _{2} \left(\mathbf{r}\right)+g_{22} \Phi _{2} \left(\mathbf{r}\right)\left|\Phi _{2} \left(\mathbf{r}\right)\right|^{2} +g_{21} \Phi _{2} \left(\mathbf{r}\right)\left|\Phi _{1} \left(\mathbf{r}\right)\right|^{2} 
&=&
\mu _{2} \Phi _{2} \left(\mathbf{r}\right).
\end{eqnarray}

\noindent The particle number for each species is defined by relation

\begin{equation} 
\label{eq2}
N_{i} =\int \left|\Phi _{i} (\mathbf{r})\right| ^{2} d\mathbf{r},\quad \quad (i=1,2). 
\end{equation}

We use the index $i$ for denoting particle species and condensate components.
Complex functions can be represented through the absolute values $\rho_{i}$ and phases $\chi_{i}$, writing them in the form of $\Phi _{i}\left(\mathbf{r}\right)=\rho_{i}\left(\mathbf{r}\right)\exp\left[i\chi_{i}\left(\mathbf{r}\right)\right]$. Then the particle number density $n_{i}\left(\mathbf{r}\right)$ and the particle velosity $\mathbf{v}_{i}\left(\mathbf{r}\right)$ are defined by the absolute value and the phase of complex function  

 \begin{equation}
\label{eq3}
n_{i} \left(\mathbf{r}\right)=\rho _{i}^{2} \left(\mathbf{r}\right),\quad \quad \mathbf{v}_{i} \left(\mathbf{r}\right)=\frac{\hbar }{m_{i} } \nabla \chi _{i} \left(\mathbf{r}\right). 
\end{equation}

\noindent In homogeneous state the chemical potentials $\mu _{1} ,\, \mu _{2} $ are 

\begin{equation}
\label{eq4}
\mu _{1} =g_{11} n_{10} +g_{12} n_{20} ,\quad \quad \mu _{2} =g_{22} n_{20} +g_{21} n_{10} . 
\end{equation}

 \noindent We assume that the stability condition for homogeneously system is satisfied~\cite{PethSm,PStr} 

\begin{equation}
\label{eq5}
g_{11} >0,\quad g_{22} >0,\quad g_{11} g_{22} -g_{12}^{2} >0. 
\end{equation}

\noindent We use the dimensionless functions for convenience: 
$\psi _{1} ={\Phi _{1} \mathord{\left/ {\vphantom {\Phi _{1}  \sqrt{n_{10} } }} \right. \kern-\nulldelimiterspace} \sqrt{n_{10} } } $, $\psi _{2} ={\Phi _{2} \mathord{\left/ {\vphantom {\Phi _{2}  \sqrt{n_{20} } }} \right. \kern-\nulldelimiterspace} \sqrt{n_{20} } }.$

\noindent The healing lengths for the first and the second condensates are as follows
\begin{equation}
\label{eq6}
\xi _{1} \equiv \frac{\hbar }{\sqrt{2m_{1} g_{11} n_{10} } } ,\quad \quad \xi _{2} \equiv \frac{\hbar }{\sqrt{2m_{2} g_{22} n_{20} } } .  
\end{equation}

\noindent Define effective healing leangth $\xi =\sqrt{\xi _{1} ^{2} +\xi _{2} ^{2} } $. Then the dimensionless coordinates will be $\tilde{\mathbf{r}}={\mathbf{r}\mathord{\left/ {\vphantom {r \xi }} \right. \kern-\nulldelimiterspace} \xi } $ and $\Delta =\xi ^{-2} \tilde{\Delta }$. Hereafter the tilde symbol will be omitted. Dimensionless form of equations~(\ref{eq1a}) and~(\ref{eq1b}) reads 

\begin{equation}
\label{eq7}
\begin{array}{c} {-\Delta \psi _{1} \left(\mathbf{r}\right)+A_{1} (\left|\psi _{1} \left(\mathbf{r}\right)\right|^{2} -1)\psi _{1} \left(\mathbf{r}\right)+uB_{1} (\left|\psi _{2} \left(\mathbf{r}\right)\right|^{2} -1)\psi _{1} \left(\mathbf{r}\right)=0,} \\
 {-\Delta \psi _{2} \left(\mathbf{r}\right)+A_{2} (\left|\psi _{2} \left(\mathbf{r}\right)\right|^{2} -1)\psi _{2} \left(\mathbf{r}\right)+uB_{2} (\left|\psi _{1} \left(\mathbf{r}\right)\right|^{2} -1)\psi _{2} \left(\mathbf{r}\right)=0.} 
\end{array}  
\end{equation}

\noindent The system of equations~(\ref{eq7}) has a symmetric form. It has five dimensionless coefficients:  $A_{1} , A_{2} , B_{1},$ $ B_{2}$ and $u\equiv {g_{12} \mathord{\left/ {\vphantom {g_{12}  \sqrt{g_{11} g_{22} } }} \right. \kern-\nulldelimiterspace} \sqrt{g_{11} g_{22} } } $.  Coefficients $A_{i} ,{\kern 1pt} {\kern 1pt} B_{i} $ are not independent. They are expressed by two dimensionless positive parameters, one of them is defined by ratio of healing lengths  $\zeta ^{2} \equiv \left({\xi _{2} \mathord{\left/ {\vphantom {\xi _{2}  \xi _{1} }} \right. \kern-\nulldelimiterspace} \xi _{1} } \right)^{2} $, and the second of them is defined by ratio of densities and interaction constants in each component  $\eta \equiv \left({n_{20} \mathord{\left/ {\vphantom {n_{20}  n_{10} }} \right. \kern-\nulldelimiterspace} n_{10} } \right)\sqrt{{g_{22} \mathord{\left/ {\vphantom {g_{22}  g_{11} }} \right. \kern-\nulldelimiterspace} g_{11} } } $.       
Thus the solution of the system of equations~(\ref{eq1a}) and~(\ref{eq1b}) depend on the following three independent dimensionless parameters  

\begin{equation}
\label{eq8}
\zeta ^{2} \equiv \left(\frac{\xi _{2} }{\xi _{1} } \right)^{2} ,\quad \quad \eta \equiv \frac{n_{20} }{n_{10} } \sqrt{\frac{g_{22} }{g_{11} } } ,\quad \quad u=\frac{g_{12} }{\sqrt{g_{11} g_{22} } } .
\end{equation}

 \noindent The first two parameters in~(\ref{eq8}) are positive but the third parameter satisfies inequality  $u^{2} <1$. The coefficients in~(\ref{eq7}) are expressed through parameters~(\ref{eq8}): 

\begin{equation}
\label{eq9}
\begin{array}{c} {A_{1} \equiv 1+\zeta ^{2} ,\quad \quad A_{2} \equiv 1+\zeta ^{-2} ,} \\ {B_{1} \equiv A_{1} \eta ,\quad \quad B_{2} \equiv A_{2} \eta ^{-1} .} \end{array}
\end{equation}

In single component case the dimensionless GP equation has not the parameters of medium and thus it is the universal equation. In two component condensate the coefficients of system are defined by three dimensionless independent parameters~(\ref{eq8}).  The vortex has different structure depending on the values of this coefficients.   

%%%%%%%%%%%%%%%%%%%%%%%%%%%%%%%%%%%%%%%%%%%%%%%%%%%%%%%%%%%%%%%%%%%%%%%%%%%%%%%%%%%%%%%%%%%%%%%%%%%%%%%

\section{Gross-Pitaevskii equation for vortex}

%%%%%%%%%%%%%%%%%%%%%%%%%%%%%%%%%%%%%%%%%%%%%%%%%%%%%%%%%%%%%%%%%%%%%%%%%%%%%%%%%%%%%%%%%%%%%%%%%%%%%%%

We seek for solution of the set of equations~(\ref{eq7}) with the symmetry about the axis of $ z $ in the form $\psi _{i} \left(r,\varphi \right)=f_{i} \left(r\right)\exp \left[i\chi _{i} \left(\varphi \right)\right]$, where $r,\varphi $ are the cilindrical coordinates. 

In this case the velocities of the components have the form

\begin{equation}
\label{eq10}
\mathbf{v}_{i} =\frac{\hbar }{m_{i} r} \frac{d\chi _{i} \left(\varphi \right)}{d\varphi } \mathbf{e}_{\varphi } ,
\end{equation}

\noindent where $\mathbf{e}_{\varphi } $ is the unit vector tangent to the circle with center on the z-axis.

The set of equations for functions $f_{i} \left(r\right)$ from~(\ref{eq7}) are: 

\begin{eqnarray}
\label{eq11a}
-\frac{d^{2} f_{1} }{dr^{2} } -\frac{1}{r} \frac{df_{1} }{dr} +\frac{l_{1} ^{2} }{r^{2} } f_{1} +A_{1} (f_{1} ^{2} -1)f_{1} +uB_{1} (f_{2} ^{2} -1)f_{1} 
&=&0,
 \\
\label{eq11b}
 -\frac{d^{2} f_{2} }{dr^{2} } -\frac{1}{r} \frac{df_{2} }{dr} +\frac{l_{2} ^{2} }{r^{2} } f_{2} +A_{2} (f_{2} ^{2} -1)f_{2} +uB_{2} (f_{1} ^{2} -1)f_{2} 
&=&
0.
\end{eqnarray}

We will study the solutions of this equations with the following boundary conditions: 

\noindent if $l_{i} ^{2} \ge 1$ we take $f_{i} \left(0\right)=0$ and $f_{i} \left(\infty \right)=1$, when $l_{i} =0$ 
we take ${df_{i}}/{dr}(0)=0$ and $f_{i} (\infty )=1$. 

The total density of the particle number in the vortex is 

\begin{equation}
\label{eq12}
n\left(r\right)=n_{10} f_{1}^{2} \left(r\right)+n_{20} f_{2}^{2} \left(r\right). 
\end{equation}

\noindent The total particle number density is equal to the equilibrium density away from the axis of the vortex
$n\left(\infty \right)\equiv n_{0} =n_{10} +n_{20} $. 

We take the phase of macroscopic wavefunction in the form
 $\chi _{i} \left(\varphi \right)=l_{i} \varphi $, where $l_{i} =0,\pm 1,\pm 2,\ldots $. The sign of $ l_{i} $ determines the direction of vortex rotation. In a two-component system the vortex motion in each component has its own quantized circulation:

\begin{equation}
\label{eq13}
\Gamma _{i} \equiv \oint \mathbf{v}_{i} \cdot d{\kern 1pt} \mathbf{l} =\frac{2\pi \hbar l_{i} }{m_{i} } . 
\end{equation}

An interesting feature of a two-component condensate is the possibility of vortex existence in one component, while the other component remains irrotational with the circulation equal to zero.
 In this case $l_{1} =\pm 1$ and $l_{2} =0$, or $l_{1} =0$ and $l_{2} =\pm 1$. 

Another significant feature is manifested in the rotatable component in opposite directions. In particular, $ l_ { 1} = + 1 $ and $ l_ { 2} = -1 $ corresponds to this case. Mass flow density in the vortex filament of two-component condensate is defined by the relation

\begin{equation}
\label{eq14}
\mathbf{J}=\frac{\hbar }{r} \left[l_{1} n_{10} f_{1}^{2} \left(r\right)+l_{2} n_{20} f_{2}^{2} \left(r\right)\right]\mathbf{e}_{\varphi } \equiv J_{\phi } \left(r\right)\mathbf{e}_{\varphi } . 
\end{equation}

\noindent With this in mind, the angular momentum of the vortex $\mathbf{L}=\int\left[\mathbf{r}\times \mathbf{J}\right]{\kern 1pt} d\mathbf{r}$ in a two-component BEC is given by

\begin{equation}
\label{eq15}
\mathbf{L}=\hbar \left(l_{1} N_{1} +l_{2} N_{2} \right)\mathbf{e}_{z} , 
\end{equation}

\noindent where $\mathbf{e}_{z} $ is a unit vector along the axis of the vortex, and $N_{i}=2\pi n_{i0} L_{{\rm v}} \int _{0}^{\infty}f_{i}^{2}\left(r\right)rdr$ is the total density of particles of species $i$, $L_{{\rm v}}$ is the length of the vortex. 
Note that the density distribution of the component does not depend on the sign of circulation, but the angular momentum~(\ref{eq15}), of course, depends on it.
If the rotation of component is the opposite, and the condition $ l_ { 1 } n_ { 10 } + l_ { 2 } n_ { 20} = 0 $ is satisfied, the total angular momentum of the vortex vanishes.

%%%%%%%%%%%%%%%%%%%%%%%%%%%%%%%%%%%%%%%%%%%%%%%%%%%%%%%%%%%%%%%%%%%%%%%%%%%%%%%%%%%%%%%%%%%%%%%%%%%%%%%%%

\section{Asymtotic solutions of equations for vortex}

%%%%%%%%%%%%%%%%%%%%%%%%%%%%%%%%%%%%%%%%%%%%%%%%%%%%%%%%%%%%%%%%%%%%%%%%%%%%%%%%%%%%%%%%%%%%%%%%%%%%%%%%%

We consider the asymptotic behavior of solutions of~(\ref{eq11a}),(\ref{eq11b}).
The solutions of this equations have the following form at large distances $ r \gg 1 $: $f_{i} \approx 1+a_{i} /r^{2} $ where 

\begin{equation}
\label{eq16}
a_{1} =-\frac{\left(l_{1}^{2} A_{2} -u{\kern 1pt} \eta l_{2}^{2} A_{1} \right)}{2A_{1} A_{2} \left(1-u^{2} \right)} ,\quad \quad a_{2} =-\frac{\left(l_{2}^{2} A_{1} -u{\kern 1pt} \eta ^{-1} l_{1}^{2} A_{2} \right)}{2A_{1} A_{2} \left(1-u^{2} \right)} . 
\end{equation}

The denominator in these formulas is always positive and the numerator can have different signs. Depending on the sign of the parameter $ a_ {i} $,the absolute value of macroscopic wave function may approach to its equilibrium value (unity) at infinity from above or from below.
Obviously, if $ u < 0 $, then both parameters $ a_ {i} $ are negative. If $ u> 0 $, then it is possible that one of these parameters is positive, and the other parameter is negative.
Suppose $l_{1}^{2} >0$, $l_{2}^{2} >0$.
Then, for $ u>0 $ the following cases can be implemented: 

1) if ${A_{2} \mathord{\left/ {\vphantom {A_{2}  A_{1} <}} \right. \kern-\nulldelimiterspace} A_{1} <} u\eta \left({l_{2}^{2} \mathord{\left/ {\vphantom {l_{2}^{2}  l_{1}^{2} }} \right. \kern-\nulldelimiterspace} l_{1}^{2} } \right)$, then $a_{1} >0,\; a_{2} <0$,  

2) if ${u\eta \left({l_{2}^{2} \mathord{\left/ {\vphantom {l_{2}^{2}  l_{1}^{2} }} \right. \kern-\nulldelimiterspace} l_{1}^{2} } \right)<A_{2} \mathord{\left/ {\vphantom {u\eta \left({l_{2}^{2} \mathord{\left/ {\vphantom {l_{2}^{2}  l_{1}^{2} }} \right. \kern-\nulldelimiterspace} l_{1}^{2} } \right)<A_{2}  A_{1} <}} \right. \kern-\nulldelimiterspace} A_{1} <} u^{-1} \eta \left({l_{2}^{2} \mathord{\left/ {\vphantom {l_{2}^{2}  l_{1}^{2} }} \right. \kern-\nulldelimiterspace} l_{1}^{2} } \right)$, then $a_{1} <0,\; a_{2} <0$, 

3) if ${A_{2} \mathord{\left/ {\vphantom {A_{2}  A_{1} >}} \right. \kern-\nulldelimiterspace} A_{1} >} u^{-1} \eta \left({l_{2}^{2} \mathord{\left/ {\vphantom {l_{2}^{2}  l_{1}^{2} }} \right. \kern-\nulldelimiterspace} l_{1}^{2} } \right)$, then $a_{1} <0,\; a_{2} >0$. 

\noindent If circulation is absent in one of the components, for example, in the second component

$l_{2} =0$, then $a_{1} ={-l_{1}^{2} A_{2} \mathord{\left/ {\vphantom {-l_{1}^{2} A_{2}  2A_{1} A_{2} \left(1-u^{2} \right)}} \right. \kern-\nulldelimiterspace} 2A_{1} A_{2} \left(1-u^{2} \right)} ,\quad a_{2} ={u\eta ^{-1} l_{1}^{2} A_{2} \mathord{\left/ {\vphantom {u\eta ^{-1} l_{1}^{2} A_{2}  2A_{1} A_{2} \left(1-u^{2} \right)}} \right. \kern-\nulldelimiterspace} 2A_{1} A_{2} \left(1-u^{2} \right)} $. 

\noindent In this case $ a_{1}<0 $ and the sign of $ a_{2} $ can be either positive or negative, and coincides with the sign of the constant of the interaction between the components
$u={g_{12} \mathord{\left/ {\vphantom {g_{12}  \sqrt{g_{11} g_{22} } }} \right. \kern-\nulldelimiterspace} \sqrt{g_{11} g_{22} } } $. 

According to~(\ref{eq12}) and~(\ref{eq16}) the behavior of the total density at large distances  defined by the formula

\begin{equation}
\label{eq17}
\frac{n\left(r\right)}{n_{0} } =1+\frac{2a_{D} }{n_{0} r^{2} } ,
\end{equation}

\noindent where

\begin{equation}
\label{eq18}
a_{D} =a_{1} n_{10} +a_{2} n_{20} =-\frac{\left[\left(n_{10} l_{1}^{2} A_{2} +n_{20} l_{2}^{2} A_{1} \right)-u\left(n_{10} \eta l_{2}^{2} A_{1} +n_{20} \eta ^{-1} l_{1}^{2} A_{2} \right)\right]}{2A_{1} A_{2} \left(1-u^{2} \right)} .   
\end{equation}

\noindent We introduce the notation 

\begin{equation}
\label{eq19}
u_{*} =\frac{\left(n_{10} l_{1}^{2} A_{2} +n_{20} l_{2}^{2} A_{1} \right)}{\left(n_{20} \eta ^{-1} l_{1}^{2} A_{2} +n_{10} \eta l_{2}^{2} A_{1} \right)}. 
\end{equation}

\noindent When $u=u_{*} $ the parameter $a_{D}$~(\ref{eq18}) is equal to zero. 
In case of positive parameter $ u_{*}>1$, for any admissible values of $ u $ we have $ a_{D} <0 $ and, therefore, the total density tends to its equilibrium value at infinity from below.
An interesting possibility arises if $ u_{*}<1 $. It is possible to have $u>u_{*}$ and $ a_{D}>0 $ in this case.
This means that the total density approaches its equilibrium value at infinity from above, i.e. density near the vortex axis is greater than the equilibrium density at infinity. We emphasize that the possibility of the existence of such ``vortices with super-density'' near the axis is a feature of the multicomponent condensation. When the circulation in the second component is absent ($l_{2} =0$), $ a_{D} $ has the simple form

\begin{equation}
\label{eq20}
a_{D} =\frac{l_{1}^{2} \left(n_{10} -u\eta ^{-1} n_{20} \right)}{2A_{1} \left(1-u^{2} \right)} .
\end{equation}

\noindent Coefficient~(\ref{eq20}) is positive if the condition $ n_{10}> u \eta^{-1} n_{20} $ is satisfied.
This condition is equivalent to $ g_{22}> g_{12} $ in accordance with~(\ref{eq8}).
In this case, the total density near the vortex axis is greater than the equilibrium value of the density at infinity.

%%%%%%%%%%%%%%%%%%%%%%%%%%%%%%%%%%%%%%%%%%%%%%%%%%%%%%%%%%%%%%%%%%%%%%%%%%%%%%%%%%%%%%%%%%%%%%%%%%%%%%%

\section{The vortex structure on the different signs of the inter-component interaction constant} 

%%%%%%%%%%%%%%%%%%%%%%%%%%%%%%%%%%%%%%%%%%%%%%%%%%%%%%%%%%%%%%%%%%%%%%%%%%%%%%%%%%%%%%%%%%%%%%%%%%%%%%%

We investigate numerically the structure of the vortex at different sign and magnitude of the parameter $u$, which determines
the ``interaction'' of BEC component.
 Consider first the case when the vortex motion in the second component is missing, so the $ l_{2} = 0 $.
Put in numerical calculations that $\zeta =\eta =1$, thus, $A_{1} =A_{2} =B_{1} =B_{2} =2$, and $u_{*} =n_{10} /n_{20} $~(\ref{eq19}).
 Inter-component interaction parameter can vary in the range $-1<u<1$. 

\begin{figure}
\includegraphics[height=73mm]{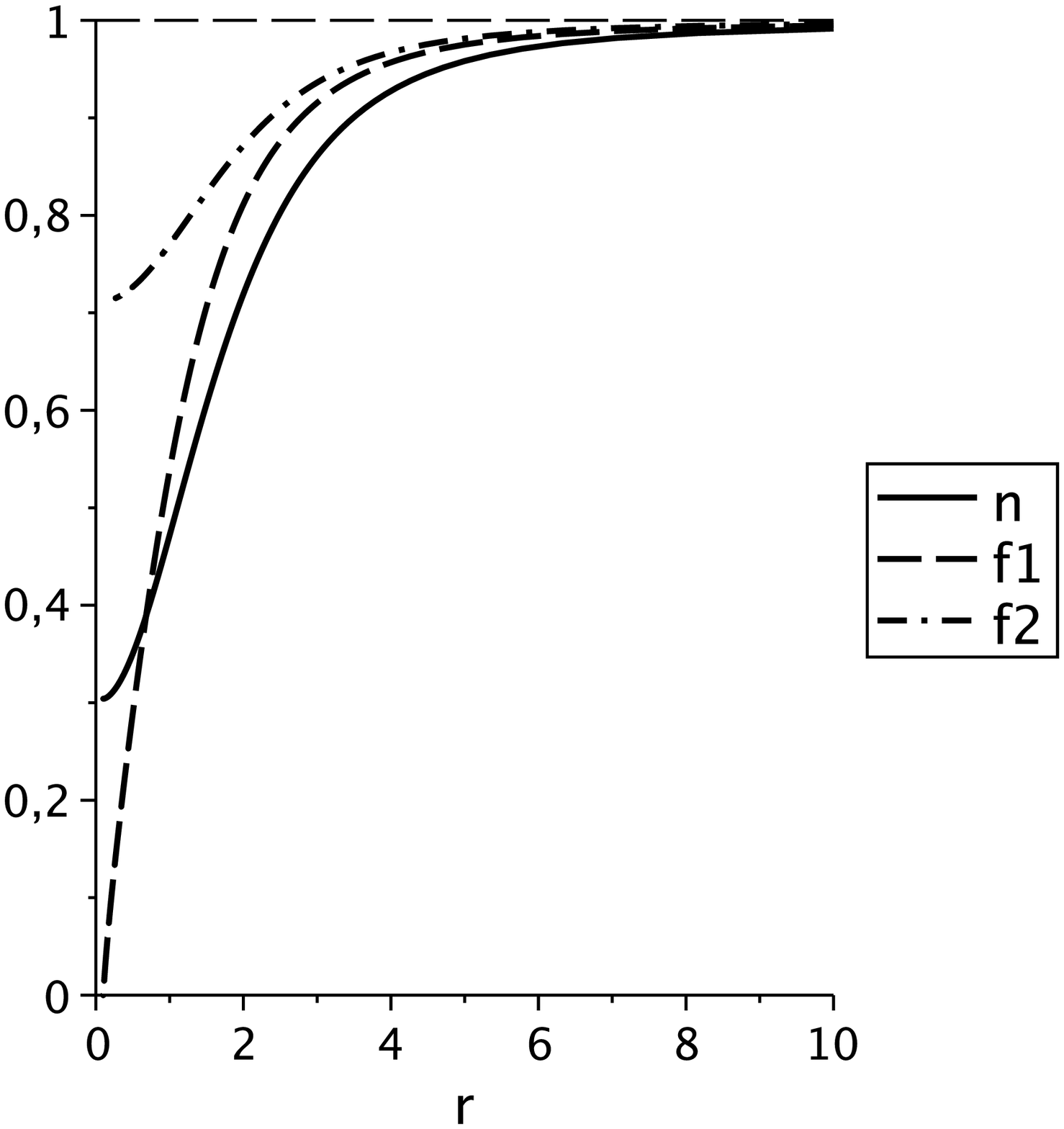} 
\hfill
\includegraphics[height=73mm]{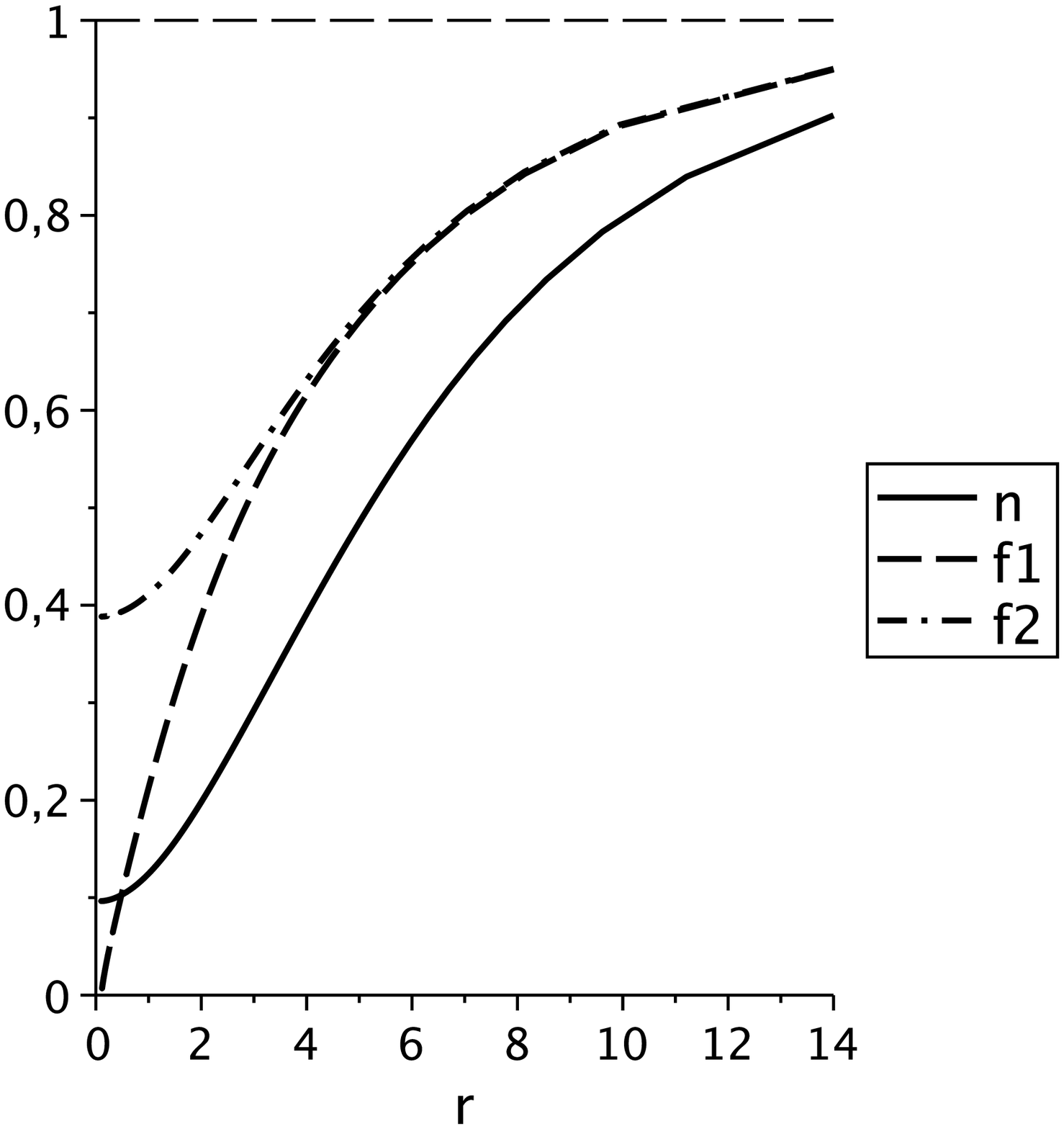} 
\begin{center}
(a)\hspace{10cm}(b)
\end{center}
\caption{
\label{fig:f1}
The solutions of GP equations for $f_{1} \left(r\right)$, $f_{2} \left(r\right)$ and for total density $n(r)$ in the case $l_{1}^{2} =1, l_{2} =0$, $u_{*} =n_{10} /n_{20} =2/3$:
a) $u=-0.7$; b) $u=-0.98$. }
\end{figure}

\begin{figure}
\includegraphics[height=73mm]{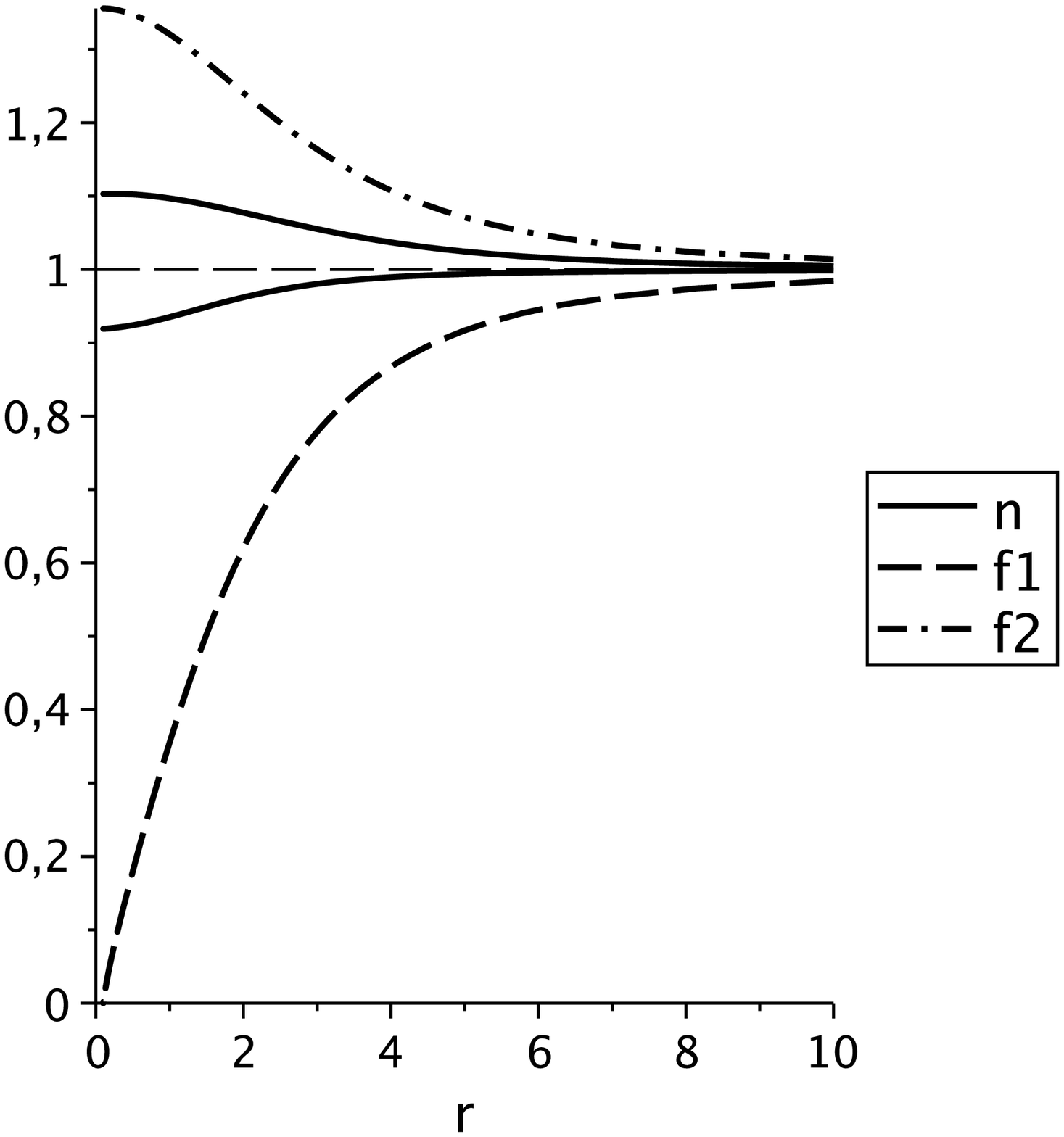} 
\hfill
\includegraphics[height=73mm]{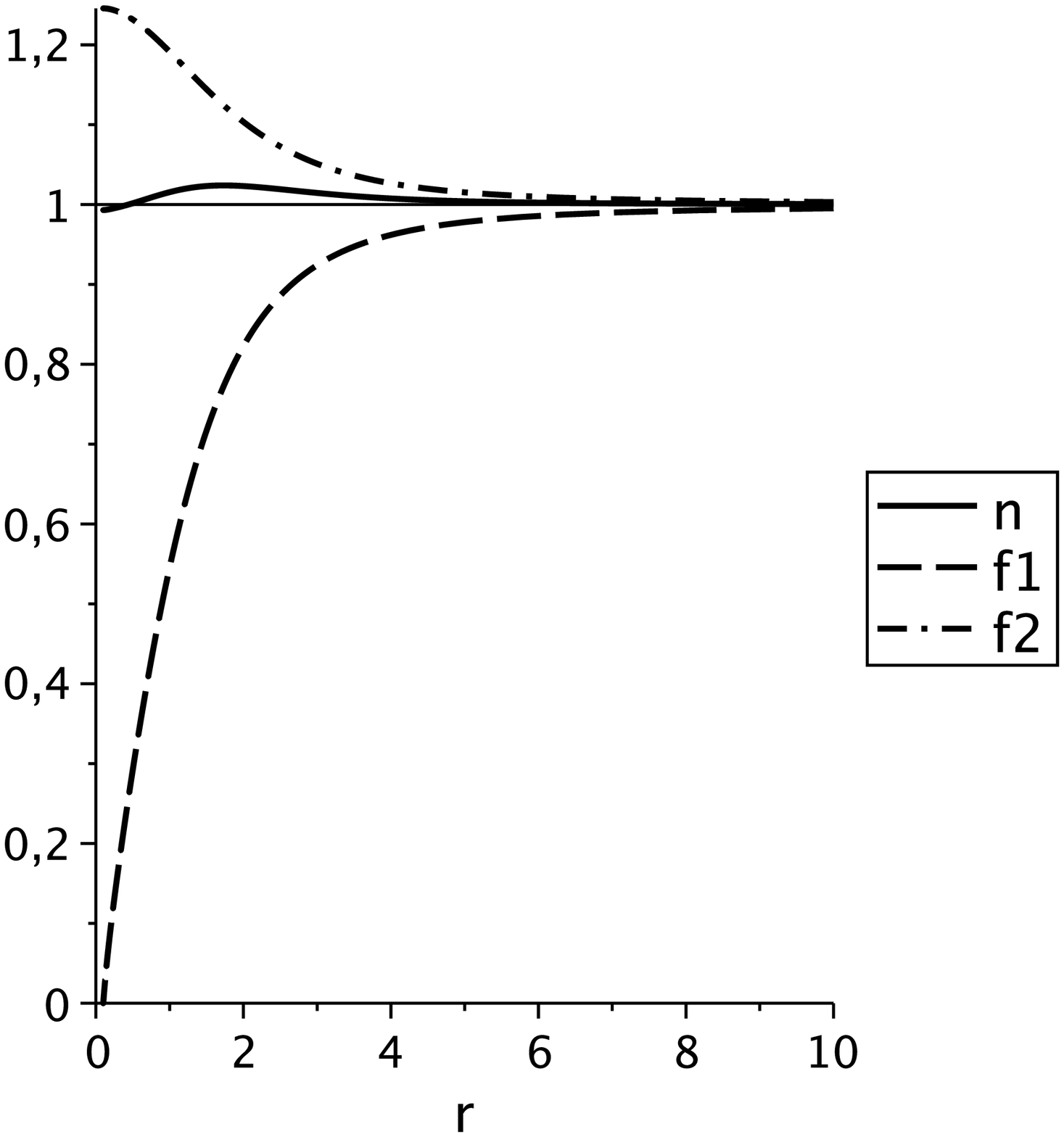} 
\begin{center}
(a)\hspace{10cm}(b)
\end{center}
\caption{
\label{fig:f2}
The solutions of GP equations for $f_{1} \left(r\right)$, $f_{2} \left(r\right)$ and for total density $n(r)$ in the case $l_{1}^{2} =1, l_{2} =0$: a) $u=0.9$; b) $u=0.6666$. The upper solid curve in Figure (a) shows the behavior of the total density $n$ if $u_{*} =n_{10} /n_{20} =2/3<0.9$. The lower curve corresponds to the parameters $u_{*} =n_{10} /n_{20} =1>0.9$. The Figure (b) shows the same function for parameters: $u=0.6666$, $u_{*}=0.5625$. }
\end{figure}

Numerical solutions of the equations~(\ref{eq11a}), (\ref{eq11b}) and the behavior of the total density as a function of distance from the vortex axis at $ u=-0.7 $, $ u=-0.98 $, $ u= 0.9 $ and $u=0.6666$ are shown in Figures~\ref{fig:f1} and~\ref{fig:f2}.
In the plots we follow the notation with dimensionless $r$.
The case $ u = -0.7 $ is shown in Figure ~\ref{fig:f1}a.
The macroscopic wave-function absolute value for component with circulation vanishes on the axis.
In component, where there is no circulation, it takes a finite value less than unity.
The wave functions of the two components and the relative total density increases monotonically with distance, approaching the equilibrium values.
The total density on the vortex axis, in contrast to the usual single-component BEC vortex, does not vanish.

Figure 1b shows a similar dependence at $ u = -0.98 $. In this case, we are near the stability limit of the system (see~(\ref{eq5})). We see that the qualitative behavior of the total density is the same as that of the single-component condensate, but the value of the total density on the axis of the vortex is nonzero.  

Figure 2a shows graphs at $ u = 0.9 $, $ u_{*} =2/3 $ and $ u_{*} =1 $. In the case $ u_{*} =2/3 $, in contrast to the previous case, the magnitude of the wave function components without circulation on the vortex axis is greater than its value at infinity, and it decreases monotonically with increasing distance, approaching unity from above.
Absolute value of the wave function components with vortex motion increases monotonically with distance from zero to the equilibrium value.

The behavior of the total density in this case may be different: if $ u> u_{*} $ it decreases monotonically from a value greater than the equilibrium value (upper solid curve in Fig.~\ref{fig:f2}a), and at $ u <u_{*} $ density increases monotonically from the smaller unit on the axis (the lower solid curve in Figure~\ref{fig:f2}a).

Some interesting features in the structure of vortices in two-component condensate
are found for certain relations between the parameters $u$ and $u_{*}$.
The case $ u=0.6666$,  $u_{*} =0.5625$ is shown in Figure~\ref{fig:f2}b.
Here the total density remains almost unperturbed, though the circulation is present in the system.

The numerically obtained vortex solutions with finite density on the axis and vortex solution in which the density on the axis greater than the equilibrium are qualitatively different from the usual vortices in one-component BEC.
Numerical calculations for a smaller absolute value of constant interaction $ \left|u\right| $ show that  the qualitative behavior of solutions of the equation GP and the total density does not change in this case.
When the interaction between the components is reduced, the distribution of the wave function in the irrotational component becomes more uniform, approaching the uniform distribution of $ f_{2} = 1 $.

Consider the case, where two circulations are associated with a vortex,
$l_{1}^{2} =l_{2}^{2} =1$. Here, as above, we believe that $\zeta =1$,
parameter $\eta $ can have any positive value.
 Thus, $A_{1} =A_{2} =2$ è $B_{1} =2\eta ,\; \; B_{2} =2\eta ^{-1} $. Inter-component interaction parameter can vary in the range $-1<u<1$. 
Here, as in the above case, there may be situations when the density of the components approach their equilibrium values at infinity from below and from above, depending on the sign of constant $a_{i}$~(\ref{eq16}). The following cases are possible:

1) if $u>\eta $ we have $a_{1} <0,\; a_{2} >0$; 

2) if $\eta >{1\mathord{\left/ {\vphantom {1 u}} \right. \kern-\nulldelimiterspace} u} $ we have $a_{1} >0,\; a_{2} <0$;
 
3) if $u<\eta <{1\mathord{\left/ {\vphantom {1 u}} \right. \kern-\nulldelimiterspace} u} $ we have $a_{1} <0,\; a_{2} <0$.

%%%%%%%%%%%%%%%%%%%%%%%%%%%%%%%%%%%%%%%%%%%%%%%%%%%%%%%%%%%%%%%%%%%%%%%%%%%%%%%%%%%%%%%%%%%%%%%%%%%%%

\begin{figure}
\includegraphics[height=73mm]{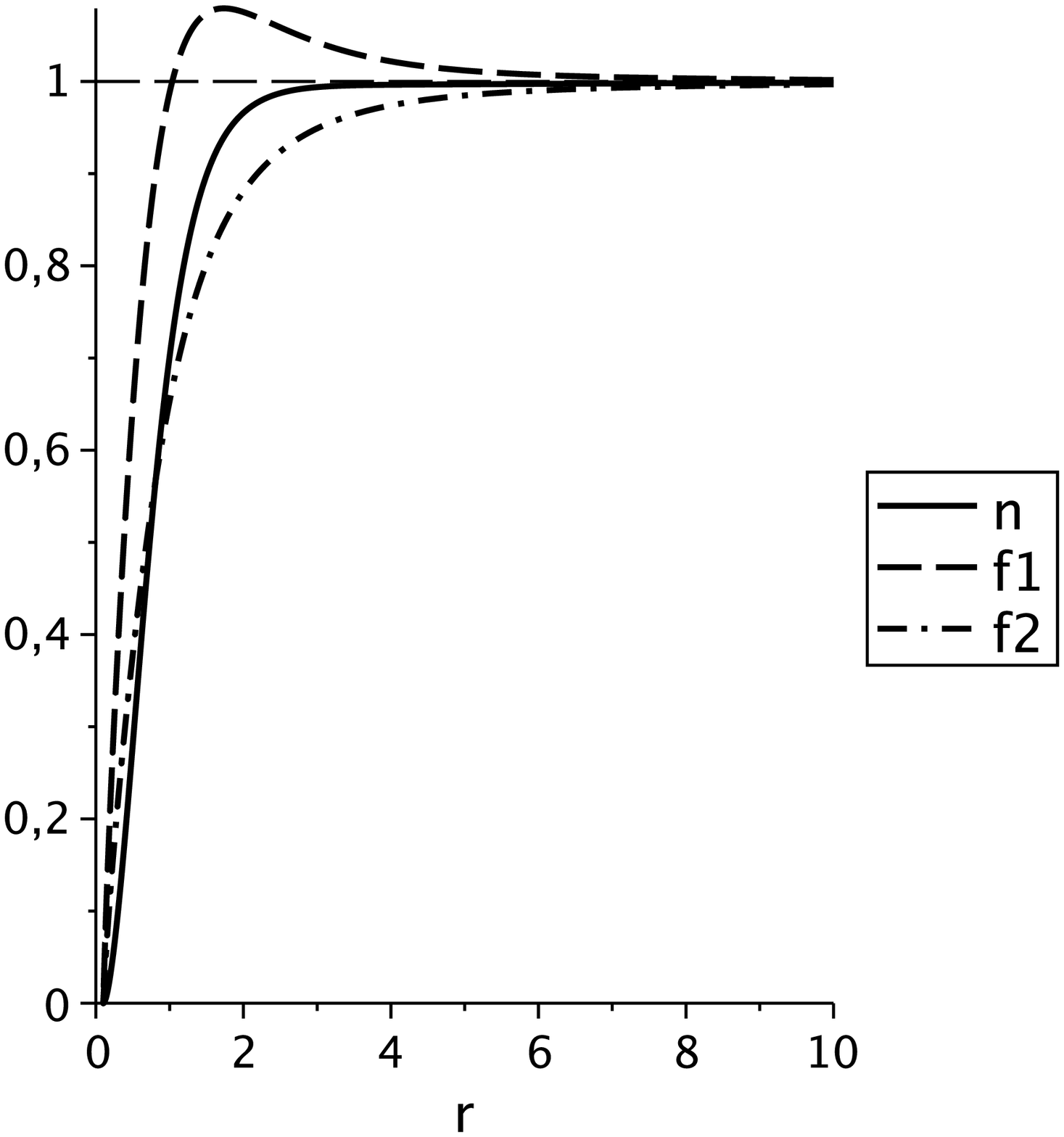} 
\hfill
\includegraphics[height=73mm]{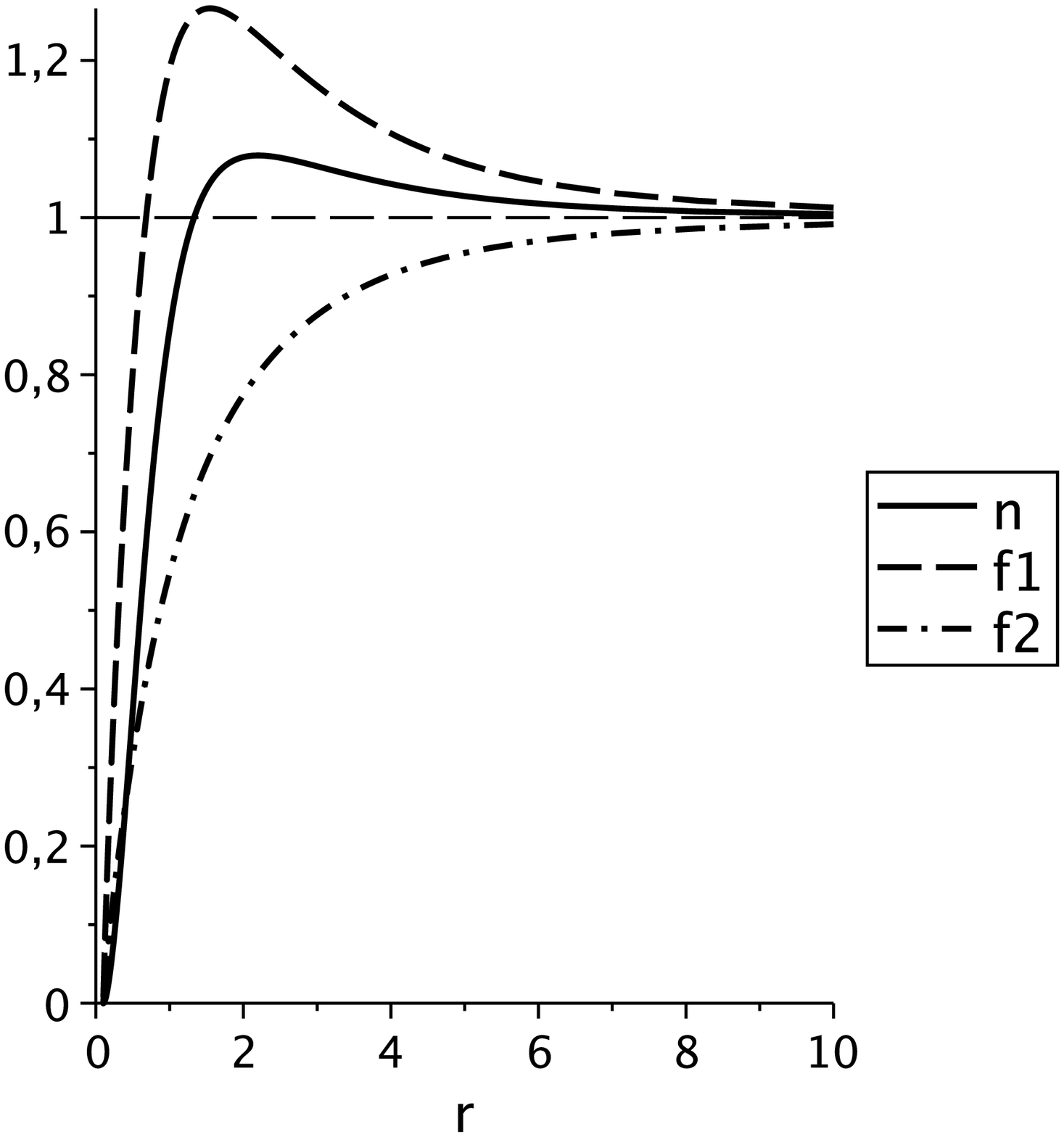} 
\begin{center}
(a)\hspace{10cm}(b)
\end{center}
\caption{
\label{fig:f3}
The solutions of GP equations for $f_{1} \left(r\right)$, $f_{2} \left(r\right)$ and for the total density $n\left(r\right)$ if $l_{1}^{2} =l_{2}^{2} =1$:
 a) $u=0.7$, b) $u=0.9$ ($\zeta =1$, $\eta =2$, $u_{*} =0.8$, $n_{10} =n_{20} $).}
 \end{figure}

\noindent Figure~\ref{fig:f3} shows the dependence of the solutions of the GP equations at the parameters, when one of the solutions tend to the value at infinity from below, and the other solution tends from above. If the parameter $ u $ is less than the critical value of $u_{*}$~(\ref{eq19}) (Fig.~\ref{fig:f3}a), the total density increases monotonically with increasing distance from the axis.

If $ u>u_{*} $, then the vortex structure is realized in a qualitatively different manner from the structure of the vortex in the one-component case (Fig.~\ref{fig:f3}b). The density increases from zero on the axis to a maximum value that exceeds the density at infinity and then monotonically decreases with the distance to equilibrium density.
In this case vortex has the form of tube
nearly empty within and limited by solid wall. 

Thus, it is shown that there may be two types of ``super-density vortices''.
In one case, the density on the axis of the vortex exceeds the equilibrium density and decreases monotonically with distance (the upper solid curve in Figure ~\ref{fig:f2}a). In another case, having a density equal to zero at the axis, the density reaches the maximum value, greater than the equilibrium density, at a certain distance of the order of the correlation length (Fig.~\ref{fig:f3}b).

%%%%%%%%%%%%%%%%%%%%%%%%%%%%%%%%%%%%%%%%%%%%%%%%%%%%%%%%%%%%%%%%%%%%%%%%%%%%%%%%%%%%%%%%%%%%%%%%%%%%%%

\section{Vortex mass flux density}

%%%%%%%%%%%%%%%%%%%%%%%%%%%%%%%%%%%%%%%%%%%%%%%%%%%%%%%%%%%%%%%%%%%%%%%%%%%%%%%%%%%%%%%%%%%%%%%%%%%%%%%

It is interesting to consider how the mass flux density varies with distance from the vortex axis. The mass flux density is defined by the formula~(\ref{eq14}), which is conveniently written in dimensionless form

\begin{equation}
\label{eq21}
\frac{J_{\phi } \left(r\right)}{J_{\phi 0} } =\frac{1}{r} \left[l_{1} \frac{n_{10} }{n_{0} } f_{1}^{2} \left(r\right)+l_{2} \frac{n_{20} }{n_{0} } f_{2}^{2} \left(r\right)\right], 
\end{equation}

\noindent where $J_{\phi 0} ={\hbar n_{0} \mathord{\left/ {\vphantom {\hbar n_{0}  \xi }} \right. \kern-\nulldelimiterspace} \xi } $. In the case $r\ll1$ è $l_{1}^{2} \ne 0,\; \; l_{2}^{2} \ne 0$ the solution of equations~(\ref{eq11a}), (\ref{eq11b}) vanish under the law $f_{1} \approx C_{1} r,\; \, f_{2} \approx C_{2} r$. Finding the expression for the mass flux density at small and large distances using the condition that $ r \gg 1 $ we have $ f_{1} \approx f_{2} \approx 1 $:

\begin{equation}
\label{eq22}
\frac{J_{\phi } \left(r\right)}{J_{\phi 0} } \approx \gamma _{0} r,\quad \left(r\ll1\right);\quad \quad \quad \frac{J_{\phi } \left(r\right)}{J_{\phi 0} } \approx \frac{\gamma _{\infty } }{r} ,\quad \left(r\gg1\right).  
\end{equation}

Direction of rotation near the vortex axis and at large distances are determined by the signs of the coefficients

\begin{equation}
\label{eq23}
\gamma _{0} \equiv l_{1} \frac{n_{10} }{n_{0} } C_{1}^{2} +l_{2} \frac{n_{20} }{n_{0} } C_{2}^{2} ,\quad \quad \gamma _{\infty } \equiv l_{1} \frac{n_{10} }{n_{0} } +l_{2} \frac{n_{20} }{n_{0} } . 
\end{equation}

\begin{figure}
\hspace{-0.5cm}
\includegraphics[height=73mm]{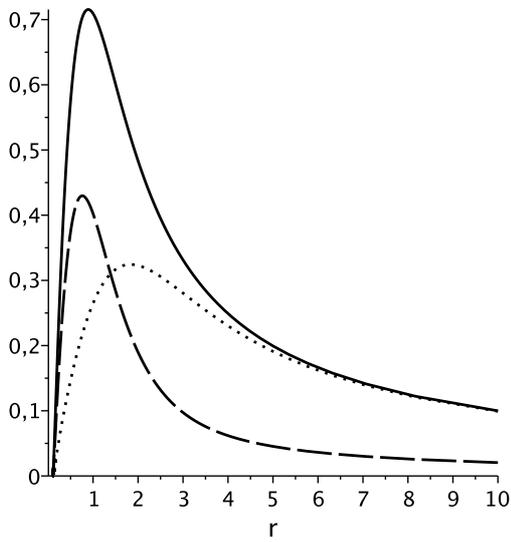} 
\hspace{2cm}
\includegraphics[height=73mm]{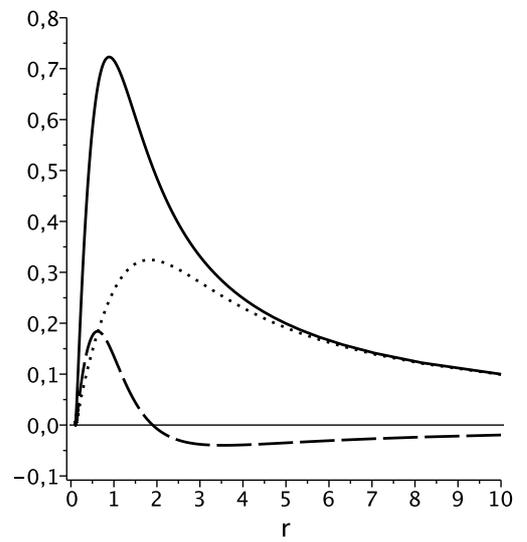} 
\begin{center}
(a)\hspace{10cm}(b)
\end{center}
\caption{
\label{fig:f4}
Full mass flux density as a function of distance from the vortex axis.
The ordinate value is 
$J_{\phi } \left(r\right)/J_{\phi 0} $,
where $J_{\phi 0} =\hbar n_{0} / \xi  $. 
The solid curves correspond to the same sign $ l_{1} $ and $ l_{2} $, and the dashed curves correspond to the different sign.
Figure (a) shows a case of the same direction of vortex rotation. Figure (b) shows the case of the direction change of vortex rotation.
The dotted curve refers to the usual one-component case.
}
\end{figure}

The coefficients $C_{1} $, $C_{2}$ can not be calculated analytically, but the numerical calculations show that the coefficients~(\ref{eq23}) may have different signs, when the signs of $ l_{1} $ and $ l_{2} $  are opposite.
This means that in this case the condensate in the vortex near the axis and at large distances is rotated  in opposite directions. Figure~\ref{fig:f4} shows the dependence of the mass flow density in the vortex of the distance from its axis.
The solid curve corresponds to the case when the signs $ l_{1} $ and $ l_{2} $ are the same ($ l_{ 1}^{2}=l_{2}^{2}=1 $). In this case, the distribution of the mass flow is not qualitatively different from the distribution in one-component BEC (dotted curve).

The most interesting case is, when the signs $ l_{1} $ and $ l_{2} $ are opposite. Dependence of the mass flow density in the vortex on the distance from the axis is shown by dashed lines in this case. The dependence on Figure ~\ref{fig:f4}a is calculated for a set of parameters: $ n_{10} /n_{20} = 3/2 $, $ \zeta =\sqrt{3/2} $, $ \eta = 4/3 $, $ u = 0.7 $.
The rotation of the condensate at any distance from the axis occurs in the same direction in this case. When parameter set is $ n_{10}/n_{20} = 2/3 $, $ \zeta =\sqrt{2/3} $, $ \eta = 3 $, $u = 0.7 $ the situation with the condensate rotation near the vortex axis and at large distances in opposite directions is  realized (Figure ~\ref{fig:f4}b).

Note that some of the effects of quantized vortices with opposite rotation in the two-component condensates in atomic gases have also been studied in a recent paper~\cite{ITT}.

%%%%%%%%%%%%%%%%%%%%%%%%%%%%%%%%%%%%%%%%%%%%%%%%%%%%%%%%%%%%%%%%%%%%%%%%%%%%%%%%%%%%%%%%%%%%%%%%%%%%%%%

\section{Conclusion}

%%%%%%%%%%%%%%%%%%%%%%%%%%%%%%%%%%%%%%%%%%%%%%%%%%%%%%%%%%%%%%%%%%%%%%%%%%%%%%%%%%%%%%%%%%%%%%%%%%%%%%%

In the one-component BEC the vortex structure is universal and does not depend on the characteristics of the medium. There is a kind of a law of corresponding states, consisting in the fact that the equation for vortex, written in dimensionless form does not contain any parameters characterizing the system.
This does not hold for the condensates, consisting from particles of different species. In particular, a two-component condensate vortices are characterized by three independent dimensionless parameters and, depending on the ratio between them, may have a different structure.

Some possible vortex structure bearing one and two quantized circulations in the two component BEC are discussed in this paper. Numerical solutions of GP equations are shown at different parameters. It is shown that the density distribution in the vortex depends on the sign of the parameter describing the interaction  between the components.
For positive parameter, there are solutions with density near the vortex axis greater than the equilibrium density at infinity (``super-density vortices'').
If the vortex has one quantum circulation, its density on the axis can reach a maximum value exceeding the equilibrium density. For vortices with two quantized circulations the density on the axis is zero and the maximum density that exceeds the equilibrium value is achieved at a distance from the axis of the correlation length order.

Another interesting feature of the vortices discussed in this paper is the ability to change the direction of rotation within the vortex. It is shown that vortices with opposite direction of rotation near the vortex axis and away from it can exist. The asymptotic behavior of vortex solutions over long distances is studied. It is shown that their analysis allows qualitative conclusions about the structure of the vortex.

\end{document}